\documentclass[%
 reprint,
superscriptaddress,
 amsmath,amssymb,
 aps,
pre,
]{revtex4-2}

\usepackage{graphics}
\usepackage{dcolumn}
\usepackage{bm}
\usepackage{makecell} 

\usepackage{color}
\usepackage{overpic}
\usepackage{xcolor}
\usepackage{soul}

\newcommand\dd{\mathrm{d}}
\newcommand\pp{\partial}

\newcommand\x{\mathbf{x}}
\newcommand\n{\mathbf{n}}

\newcommand\Q{\mathbf{Q}}

\begin{document}

\preprint{APS/123-QED}

\title{Quantum-Dots Dispersed Bent-core Nematic Liquid Crystal and Cybotactic Clusters: Experimental and Theoretical Insights}

\author{Sourav Patranabish}
\affiliation{
Department of Physics, Indian Institute of Technology Delhi, Hauz Khas, New Delhi 110016, India
}
 
\author{Yiwei Wang}%
\affiliation{
Department of Applied Mathematics, Illinois Institute of Technology, Chicago, IL 60616, USA}


\author{Aloka Sinha}
\affiliation{
Department of Physics, Indian Institute of Technology Delhi, Hauz Khas, New Delhi 110016, India
}

\author{Apala Majumdar}
\affiliation{Department of Mathematics and Statistics, University of Strathclyde, Glasgow, United Kingdom}

\begin{abstract}
We study a quantum-dots (QDs) dispersed bent-core nematic liquid crystalline system in planar geometry and present experimental measurements of the birefringence ($\Delta n$), order parameter ($S$), dielectric dispersion and absorption spectra, optical textures, with attention to variations with temperature. A bent-core liquid crystal (LC) 14-2M-CH$_3$ is used as the host material and CdSe/ZnS core-shell type QDs are used as the dopant. The nematic (N) phase of the pristine (undoped) LC 14-2M-CH$_3$ contains cybotactic clusters, which are retained by its QDs incorporated LC nanocomposite. Our experimental findings support (i) reduced orientational order parameter of the QDs dispersed LC system compared to its pristine counterpart, at fixed temperatures, (ii) reduced cybotactic cluster sizes due to the incorporation of QDs and (iii) increased activation energies related to reduced cluster sizes. We complement the experiments with a novel Landau-de Gennes type free energy for a doped bent core LC system, that qualitatively captures the doping-induced reduced order parameter and its dependence on the properties of the QDs, and its variation with temperature. 

\end{abstract}

\maketitle


\section{Introduction}

The bent-core liquid crystals (BLCs) are a novel class of liquid crystal (LC) mesogens that manifest various unique and exciting properties such as chirality, ferroelectricity and biaxiality \cite{Photinos_biaxial_JMC,Takezoe_BLC_JJAP,Jakli_BLC_LCR,Francescangeli_cybo_SM,Punjani_Golam,Keith_NBLC_SM}. They are known to form several exotic mesophases such as the twist-bend nematic (N$_{tb}$) phase, the blue phase (BP) and the banana (B1-B7) phases \cite{Takezoe_BLC_JJAP,Cestari_Ntb_PRE,V_Borshch_Ntb_NatCom,Jakli_BLC_doped_JMC}. The nematic (N) phase of BLCs itself manifests a few of the aforementioned distinct features, such as ferroelectric response, fast switching and macroscopic biaxiality \cite{Shankar_Cybo_CPC,Shankar_Cybo_AFM,Ghosh_BLC_Ferro_JMCC,Francescangeli_Ferro_AFM,Photinos_biaxial_JMC,Francescangeli_cybo_SM}. The main reason behind these extraordinary features is the locally polar cybotactic clusters formed by BLC molecules in their N phase \cite{Francescangeli_cybo_SM,Punjani_Golam,Keith_NBLC_SM, Shankar_Cybo_CPC,Shankar_Cybo_AFM,Ghosh_BLC_Ferro_JMCC,Francescangeli_Ferro_AFM}. Due to bent molecular shape and the lack of translational symmetry, the BLC molecules in their N phase experience steric hindrance. This causes stacking of the BLC molecules in smectic layers (clusters) \cite{Scaramuzza_BLC_JAP,Jakli_Rheological_SM}. These stacks of molecules are termed as `cybotactic' clusters because they are more ordered compared to the surrounding molecules. The clusters and the other BLC molecules together constitute the macroscopic N phase \cite{Francescangeli_cybo_SM}. Recent reports, aided by various experimental techniques, have established the existence of cybotactic clusters in the nematic, smectic and even in the isotropic phases \cite{Kashima_PolarBLC_JMCC,Alaasar_Cluster_JMCC,Ghosh_BLC_Ferro_JMCC,Jakli_BLC_mixture_PRE,Domenici_NMR_SM,Goodby_Unusual_SM}. Although studied extensively, the origins of cluster formation and the effects of external factors (e.g. nanoparticle doping, electric field) on these clusters remain an open problem. Further studies are required for the manipulation and successful tailoring of cybotactic clusters for applications in science and technology, including novel BLC-driven devices. \\

Suspension of nanoparticles (NPs) in the LC matrix to improve or to selectively modify the physical properties of LCs is a widely used technique in today’s liquid crystal science. Studies have shown that the dispersion of nanoparticles in LCs can improve the electro-optic properties, modify the elastic anisotropy and the dielectric constants, and reduce the transition temperatures \cite{Takatoh_LCNP_JJAP, NAClark_LCCNT_APL, WLee_TNLC_APL, Ghosh_BLCCNT_JML, JitendraK_QDBLC_JML}. The incorporation of NPs can also affect the orientation of LCs and induce a homeotropic alignment \cite{Hegmann_NPLC_LC}. Varying the size and shapes of the dopant NPs also have a profound effect on the physical properties of LCs \cite{Orlandi_LCNP_PCCP, Mirzaei_LCQD_composite_JMC, Kinkead_QDLC_JMC}. Recently, a new class of semiconductor NPs have been discovered, called the quantum dots (QDs). Incorporation of these QDs in the LC matrix may also affect or alter the physical properties of LCs, such as a reduction in the dielectric anisotropy, faster response times, changes in the phase transition temperatures and altered boundary conditions \cite{Mirzaei_LCQD_composite_JMC, Kinkead_QDLC_JMC,Mirzaei_QDLC_dopant_JMC,Zhang_LCQD_JJAP,Urbanski_NPLC_Bulk_CPC,JitendraK_QDBLC_JML}. Changes in the dielectric anisotropy ($\Delta\epsilon$) provide an indirect measure of changes in the order parameter ($S$), because $\Delta\epsilon \propto S$ \cite{JitendraK_QDBLC_JML, maier_orderparameter}. The QDs are usually capped with functionalized ligands that prevent aggregation. In particular, this makes QDs good candidates for stabilising dilute  suspensions in doping or dispersion LC experiments. To date, there has been work on QDs dispersed in calamitic nematic LCs (NLCs) while their effect on bent-core NLCs is relatively open \cite{JitendraK_QDBLC_JML}. In particular, little is known about the effect of QDs or doping in general, on the cybotactic clusters in bent core NLCs and in the absence of systematic experimental and theoretical studies on these lines, doped bent core NLC systems cannot meet their full potential. \\

We study a dilute homogeneous suspension of a QD-doped thermotropic BLC (details in the next section), confined in a planar cell with fixed boundary conditions on both cell surfaces. In particular, the undoped counterpart exhibits cybotactic clusters. Our primary investigations concern comparisons between the doped and undoped system, that give quantitative and qualitative insight into the effects of doping, the interplay between doping and cluster formation and how these effects can be tailored by temperature and external stimuli. This paper builds on our first paper \cite{patranabish2019one} wherein we focussed on a one-dimensional theoretical study of the N phase of a BLC, confined in a planar cell, within a phenomenological Landau-de Gennes (LdG) framework inspired by previous insightful modelling in \cite{madhusudana2017two}. This model is based on the premise that the N phase of BLC is characterized by two order parameters: $S_g$ that measures the ordering of the ground-state molecules (outside the clusters) and $S_c$ that measures the ordering within the smectic-like cybotactic clusters, with coupling between the two effects captured by an empirical parameter $\gamma$. In \cite{patranabish2019one}, we theoretically study the effects of spatial inhomogeneities, confinement and the coupling parameter, $\gamma$, on $S_g$ and $S_c$. Little is known about the material-dependent values of $\gamma$ or indeed how it could be experimentally controlled. Our theoretical studies showed that larger values of $\gamma$ substantially increase the interior values of $S_g$ and $S_c$ i.e. $\gamma$ promotes order outside and within the clusters of the N phase of the BLC, the effects being more pronounced for lower temperatures. However, the coupling also enhances the values of $S_g$, for temperatures above the nematic-isotropic transition temperature i.e. the bent core NLC can exhibit nematic order when the calamitic N phase does not e.g. for temperatures above the nematic-isotropic transition temperature. The model in \cite{patranabish2019one} is simplified in many ways, but yet sheds qualitative insight into the powerful prospects offered by cybotactic clusters in BLCs and how they can be used to manipulate nematic order and phase transitions for tailor-made applications.\\ 

In this paper, we report a combined experimental and theoretical analysis of a QDs dispersed bent-core nematic LC 14-2M-CH$_3$, in the dilute regime. The dilute regime applies to systems of nano-scale QDs (much smaller than the system size) with a low concentration of QDs, and the QDs are uniformly dispersed without any aggregation effects. We perform optical texture observations, dielectric measurements, optical birefringence measurements and the orientational order parameter calculations on the pristine BLC and its QDs-dispersed counterpart. The N phase of 14-2M-CH$_3$ contains cybotactic clusters, as already reported in our earlier work \cite{Patranabish_JML}. We find that the N phase of the QDs-dispersed counterpart also contains cybotactic clusters, albeit with modified properties. We report a number of interesting experimental results for the QDs-dispersed BLC system - the optical birefringence ($\Delta n$) is lowered and the macroscopic order parameter ($S$) is reduced compared to the undoped counterpart for a given temperature, the activation energy ($E_a$) increases compared to the undoped counterpart and based on the measurements of the relaxation frequencies ($f_R$) and activation energies, we deduce that the size of the cybotactic clusters decreases with QDs doping. We complement our experiments with a theoretical LdG-type model for the N phase of the QD-doped BLC, using the framework developed in \cite{canevari2019design}. This framework is not specific to QDs or to BLCs but to generic dilute doped LC systems and effectively captures the effects of the homogeneously suspended inclusions (in this case QDs) in terms of an additional contribution to the free energy. Hence, we apply this approach to the LdG free energy of a BLC system proposed in \cite{patranabish2019one} and \cite{madhusudana2017two} and qualitatively capture the effects of the QDs by means of suitable novel additional energetic terms. These additional terms, in principle, depend on the properties of the QDs e.g. size, shape, anchoring and preferred order etc. We introduce a weighted mean scalar order parameter, $S_m$, the theoretical analogue of the experimentally measured scalar order parameter. This simplistic approach does capture the doping-induced reduction in the mean order parameter $S_m$, which in turn qualitatively explains the reduction in birefringence, dielectric anisotropy. We present our experimental results in three parts below, followed by the mathematical model, numerical results and perspectives for future work.

\section{Experimental}

\begin{table}[b]
\caption{Phase sequence and transition temperatures observed in this study (using POM) during slow cooling.}
\begin{ruledtabular}
\begin{tabular}{lc} 

Compound & \makecell{Phase sequence and transition \\ temperatures ($^\circ$C)}\\
  \hline
  14-2M-CH$_3$ & Iso 134 N$_{Cyb}$ 106 Cryst. \\ 
 
  14-2M-CH$_3$ + 0.5 wt\% QDs & Iso 134 N$_{Cyb}$ 104 Cryst. \\ 
 
\end{tabular}
\end{ruledtabular}
\end{table}

A thermotropic bent-core nematic liquid crystal (LC) 14-2M-CH$_3$ was used for the experimental study and also as the host for the studied LC nanocomposite. The LC material was obtained from Prof. N.V.S. Rao's group at the Department of Chemistry, Assam University, Silchar, Assam, India. The molecular formula of 14-2M-CH$_3$, synthetic scheme details etc. are available in our earlier paper \cite{Patranabish_JML}. The CdSe/ZnS core-shell type quantum dots (QDs) of diameter 5.6 nm (Core dia: 2.8 nm + Shell thickness 1.4 nm) were procured from Sigma-Aldrich, Merck (USA) for preparing the LC nanocomposites. The spherical QDs were stabilized with the encapsulation of octadecylamine ligands, have absorption maxima in the range from 510 to 540 nm and emission wavelengths lying in the range of 530 to 540 nm, as provided by the manufacturer. The sequence of performed experimental steps are as follows: preparation of QDs dispersed LC nanocomposite, optical texture observation and evaluation of transition temperatures, orientational order parameter determination $via$ optical birefringence measurements, and dielectric characterization. All the experimental measurements were carried out while slowly cooling the sample from the isotropic liquid. \\

\begin{figure}[t]
  \centering
  \includegraphics[width = 0.9 \linewidth]{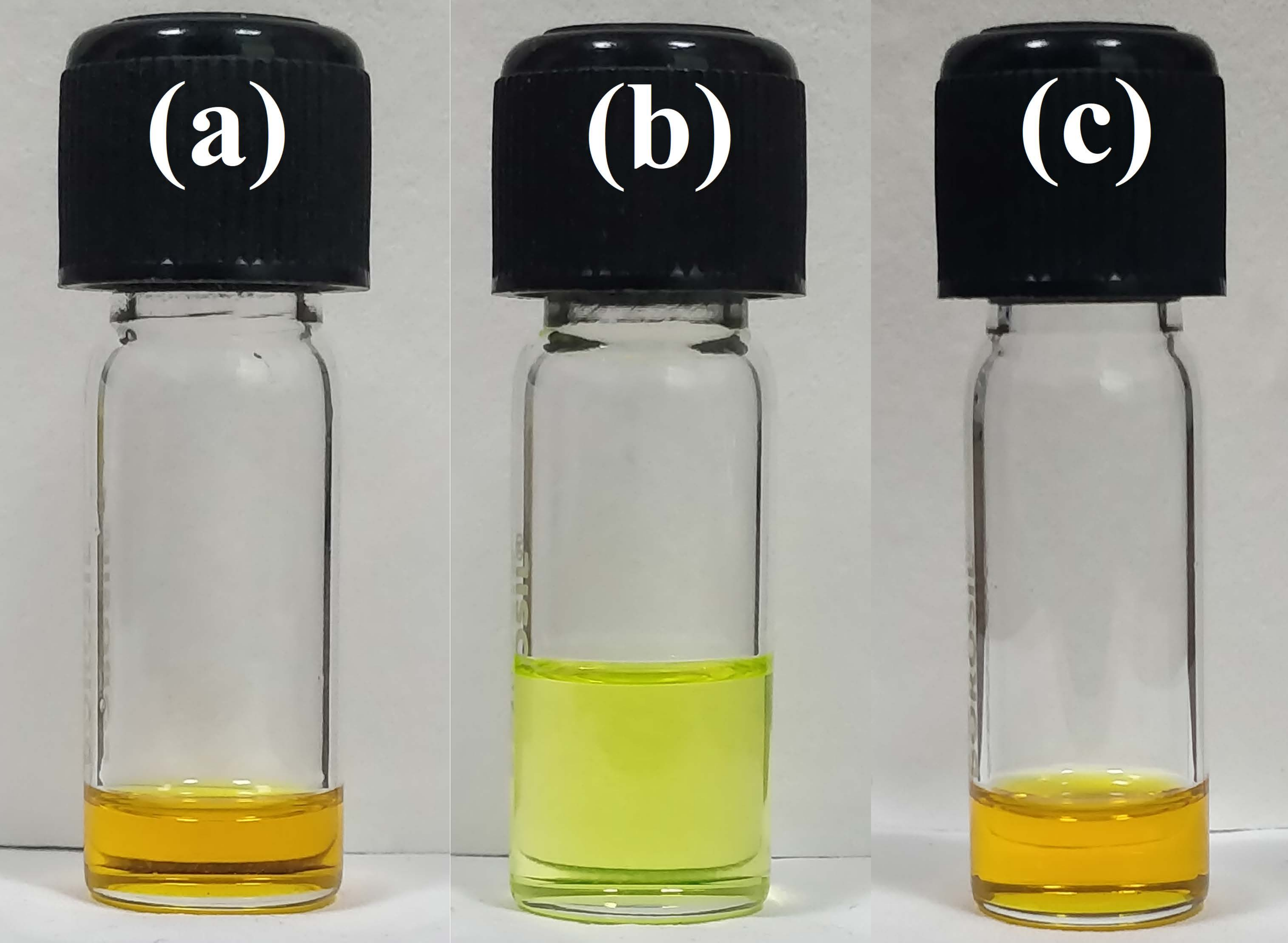}
  \caption{Visibly homogeneous solutions of (a) 14-2M-CH$_3$ (b) CdSe/ZnS QDs and (c) nanocomposite (14-2M-CH$_3$ + 0.5 wt\% CdSe/ZnS QD) in Chloroform (CHCl$_3$).}
  \label{Fig1}
\end{figure}

To prepare the LC nanocomposite, CdSe/ZnS QDs were taken at 0.5 wt\% concentration and mixed with the LC compound 14-2M-CH$_3$. To obtain a homogeneous dispersion of the quantum dots in the LC matrix, chloroform was added to the mixture, and the mixture was ultrasonicated till a visibly homogeneous dispersion was achieved (Figure 1). The mixture was kept at $\sim$ 60 $^\circ$C for 2-3 hours and it was then left overnight at room temperature for the slow evaporation of chloroform \cite{pradeepkumar}. Once the chloroform was completely evaporated, 0.5wt\% QDs dispersed LC nanocomposites were obtained. They were checked visually through a polarizing optical microscope several times, but no aggregation of QDs were noticed.\\

\begin{figure*}
\centering
\includegraphics[width = \linewidth]{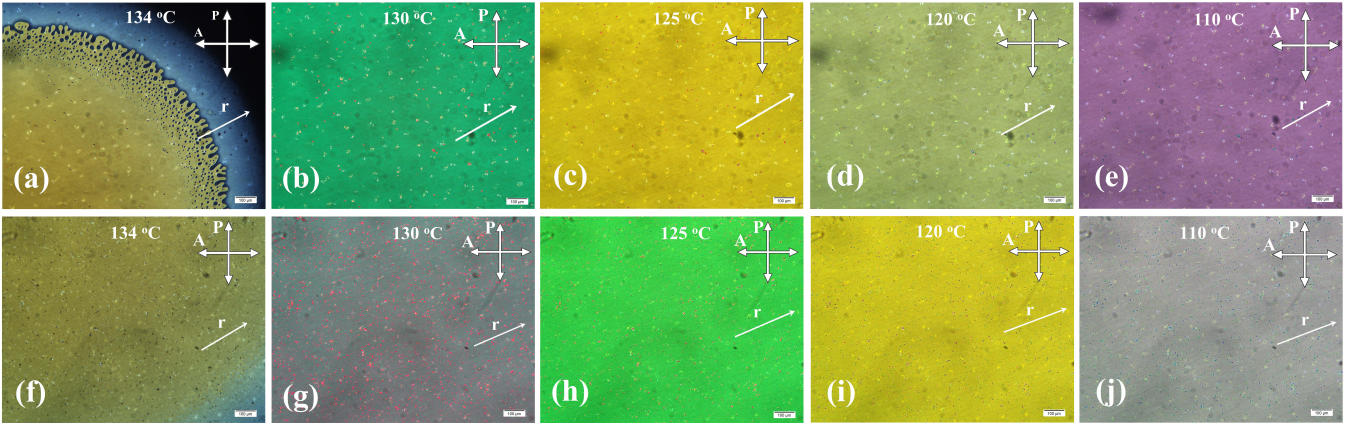}
\caption{Birefringent textural colour variation with temperature of (a-e) the bent-core LC 14-2M-CH$_3$ and (f-j) the 0.5wt\% CdSe/ZnS QDs dispersed 14-2M-CH$_3$, during cooling, respectively. In each image, \textbf{\textit{r}} indicates the rubbing direction and the scale-bar indicates 100 $\mu \mathrm{m}$. The periodic white spots in the background of the images are features of the LC cell (Instec Inc., USA) caused by PI printing fabric in the production line.}
\label{Fig2}
\end{figure*}

Indium Tin Oxide (ITO) coated 5 $\mu$m planar (homogeneous alignment) cells (Instec Inc., USA) were used for the experiments. Two different cells, of this type, were used for the pristine LC and the LC nanocomposite, respectively. The LCs were filled in the cells $via$ capillary action around 10 $^\circ$C above the clearing temperature. During measurements, the cells were kept inside an Instec HCS302 hot-stage and the temperature was maintained using an Instec MK1000 temperature controller with an accuracy of $\pm$ 0.01 $^{\circ}$C. The liquid crystalline textures were recorded using an OLYMPUS BX-51P polarizing optical microscope (POM) attached to a computer, with the sample placed between two crossed polarizers. \\

The phase behaviour and transition temperatures of the LC 14-2M-CH$_3$ and its nanocomposite were determined using the POM while slowly cooling from the isotropic liquid (0.5 $^\circ$C/min) \cite{JitendraK_QDBLC_JML, pradeepkumar}. The transition temperatures of the pristine bent-core LC 14-2M-CH$_3$ were also determined previously using differential scanning calorimetry (DSC) at a scan rate of 5 $^\circ$C/min (reported elsewhere) \cite{Patranabish_JML}. The transition temperatures of the pristine LC and its nanocomposite, as obtained from the POM observations, are summarized in Table 1. The dielectric measurements were carried out in the frequency range of 20 Hz - 2 MHz using an Agilent E4980A precision LCR meter. The measuring voltage was kept at $V_{rms} = 0.2$ V. For transmission dependent birefringence measurements and the related order parameter calculations, the sample was placed between two crossed Glan-Thompson polarizers (GTH10M, Thorlabs, Inc.) and perpendicularly illuminated with a He-Ne Laser ($\sim$ 633 nm) \cite{susantaPRE, susantaJMC}. The rubbing direction $\vec{r}$ (\textit{i.e.} the LC director $\widehat{n}$) of the planar LC cell was kept at 45$^\circ$ with respect to the polarizer (P)/analyzer (A) pass-axes. Transmitted power at the output end was measured using a Gentec PH100-Si-HA-OD1 photo-detector attached to a Gentec Maestro power meter.   \\

\section{Result and discussion}

\subsection{Polarizing optical microscopy}

The LC material is introduced in a 5 $\mu$m planar LC cell $via$ capillary action around 10 $^\circ$C above the isotropic-nematic transition temperature, and the textures were recorded between crossed polarizers. The textures recorded for the LC 14-2M-CH$_3$ and its 0.5 wt$\%$ QDs dispersed nanocomposite, during slow cooling from the isotropic liquid, are shown in Figure 2. The textures of the LC nanocomposite exhibited fairly homogeneous colours (and hence, alignment) similar to that of the pristine LC. This indicates a good, homogeneous dispersion of QDs in the LC matrix without any aggregation \cite{JitendraK_QDBLC_JML}. Close to the isotropic-nematic (Iso-N) transition temperature, we observe a sharp colour change owing to the development of nematic order in these systems (see Figures 2(a) and 2(f)). As the temperature is further lowered, uniform marble textures, typical of the nematic phase, appear with colours varying with temperature \cite{Patranabish_JML}. The isotropic-nematic transition temperature remains nearly unaltered after the incorporation of QDs. In the N phase, the emergent colours change with decreasing temperature, which indicates that the birefringence ($\Delta n$) also changes with temperature. A qualitative measurement of this change in birefringence can be made by matching the colours with the Michel-Levy chart for a given thickness \cite{Michel_Levy}. We deduce that $\Delta n$ increases with decreasing temperature from this mapping. Also, the change in $\Delta n$ with temperature is found to be quite high ($\sim$ 0.06). This is suggestive of highly ordered microstructures in the N phase of the BLC compound \cite{Keith_NBLC_SM,Nafees_RSCAdv_2015}. Also, from Figure 2 we can clearly see that the temperature dependent textural colour sequence changes/shifts after incorporation of the QDs. With the help of Michel-Levy chart, we qualitatively deduce, that the $\Delta n$ values, for a fixed temperature, are lowered on incorporation of the QDs, implying a reduction in the corresponding nematic order parameter $S$, since, $\Delta n \propto S$ \cite{pradeepkumar}. Experimentally, $\Delta n$ measurements and the associated order parameter ($S$) calculations have also been performed and they are discussed in detail in the sub-section III-B.

\subsection{Optical birefringence measurements and orientational order parameter calculations}

The birefringence ($\Delta n$) measurements of the LC sample and its nanocomposite, as a function of temperature, has been performed with the optical transmission technique. The planar LC sample is perpendicularly illuminated with a He-Ne laser ($\lambda$ $\sim$ 633 nm) and placed between two crossed Glan-Thompson polarizers such that the optic axis makes an angle $\varphi = 45^{\circ}$ with the polarizer/analyzer pass axis. The power at the output end is measured with a photodetector. The transmitted light intensity is then given in terms of the phase retardation ($\delta$) as \cite{dierking_LCtextures},

\begin{equation}
I = I_0  \sin^2 2 \varphi \sin^2  \frac{\delta}{2}   =  \frac{I_0}{2} (1 - \cos \delta),
\end{equation}

\begin{figure}[b]
    \centering
    \includegraphics[width = 0.8\linewidth]{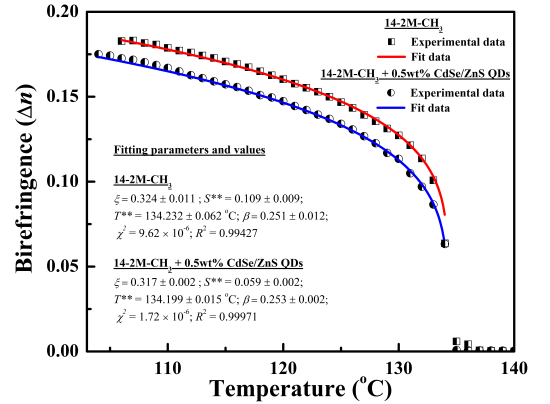}
    \caption{Experimental values of birefringence ($\Delta n$) for the LC 14-2M-CH$_3$ (half-filled squares) and its nanocomposite (half-filled circles); the solid lines (pure LC: red, LC nanocomposite: blue) represent the four-parameter fit to the experimental data using Equation (3). The related fitting parameter values are shown in the figure. The fitting parameters $\chi^2$ and $R^2$ are generated by the fitting algorithm, so that $\chi^2 \sim 0$ and $R^2 \sim 1$ describe good fits.}
    \label{Fig3}
\end{figure}

Here,  $\delta = \frac{2 \pi }{\lambda} \Delta n d$ is the phase retardation,
$I$ is the transmitted light intensity, $I_0$ is the incident light intensity, $\varphi$ (= 45$^\circ$) is the azimuthal angle, i.e., the angle made by optic axis with the polarizer/analyzer pass axis, $\lambda$ is the incident light wavelength, $\Delta n = n_e - n_o$ is the birefringence, $n_e$ and $n_o$ are the extraordinary and ordinary refractive indices of the LC, respectively, and $d$ is the thickness of the LC cell. The birefringence, $\Delta n$, is measured directly from the experimental results using equation (1), as a function of temperature. In Figure 3, we plot the experimentally measured birefringence ($\Delta n$) values for pure 14-2M-CH$_3$ (half-filled squares) and its nanocomposite (half-filled circles), at different temperatures. For both cases, on cooling from the isotropic liquid, $\Delta n$ manifests a sharp increase following the Isotropic-N phase transition, basically due to an enhancement in the nematic order. On further cooling, $\Delta n$ retains the trend but now the increase is relatively slow. It is to be noted that the birefringence values decrease appreciably due to the incorporation of QDs in the entire mesophase range.\\

\begin{figure}[t]
    \centering
    \includegraphics[width = 0.8\linewidth]{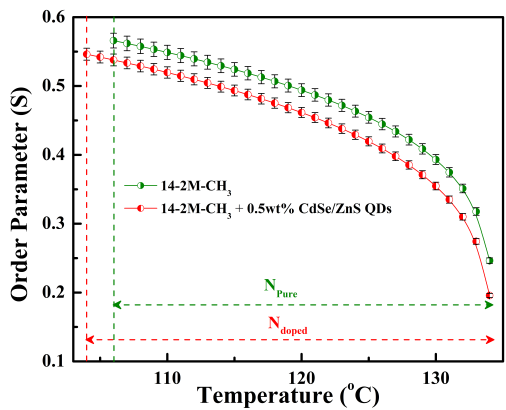}
    \caption{Orientational order parameter ($S$) as a function of temperature for the bent-core LC 14-2M-CH$_3$ and its 0.5wt\% CdSe/ZnS QDs incorporated nanocomposite.}
    \label{Fig4}
\end{figure}

For precise determination of the temperature dependence of the nematic order parameter (\textit{S}), we resort to the four-parameter power-law expression, which is in agreement with the mean-field theory of weakly first-order transitions \cite{four_parameter,susantaPRE},
\begin{equation}
S(T) = S^{**} + A \left\lvert\left(1 - \frac{T}{T^{**}}\right)\right\rvert^\beta,
\end{equation}

Here, $T$ is the absolute temperature, $T^{**}$ is the absolute superheating limit of the nematic phase; at $T=T^{**}$, $S(T^{**})=S^{**}$, $\beta$ is the critical exponent and $A$ is a constant. At $T=0$, $S(0)=1$, which implies $1 = S^{**}+A$. The birefringence, ($\Delta n$), can then be expressed as \cite{susantaPRE},
\begin{equation}
\Delta n = \xi\left[S^{**} + (1-S^{**}) \left\lvert\left(1 - \frac{T}{T^{**}}\right)\right\rvert^\beta\right],
\end{equation}

where, $\xi=(\Delta\alpha/\langle\alpha\rangle)[(n_I^2-1)/2n_I]$, $\Delta\alpha$ is the molecular polarizability anisotropy, $\langle\alpha\rangle$ is the mean polarizability and $n_I$ is the refractive index in isotropic phase just above the Isotropic-N transition temperature. The experimental birefringence ($\Delta n$) data has been well fitted with equation (3), which involves four fit parameters $\xi$, $S^{**}$, $\beta$ and $T^{**}$. The obtained fitting plots (pure LC: red solid line, LC nanocomposite: blue solid line) along with the fit parameter values, are shown in Figure 3. The four-parameter fitting is considered superior to the Haller's method that involves lesser number of fit parameters \cite{Haller_HallerFit,susantaPRE,four_parameter}. We obtain $\xi = 0.324$, $S^{**} = 0.109$, $T^{**} = 134.232 ^{\circ}C$ and $\beta = 0.251$ for the pure LC. For the LC nanocomposite, we obtain $\xi = 0.317$, $S^{**} = 0.059$, $T^{**} = 134.199 ^{\circ}C$ and $\beta = 0.253$. The fit parameter values remain almost unaltered after the incorporation of QDs, except the value of $S^{**}$, which is reduced almost by a factor of $\frac{1}{2}$. It is indicative that the QDs have a significant effect on the nematic order in the LC mesophase. The value of the critical exponent $\beta$ is around 0.25, in both cases, which is in excellent agreement with the theoretically predicted values for the nematic phase \cite{susantaPRE,four_parameter}. The temperature dependent macroscopic orientational order parameter ($S$) is calculated using equation (2) and using the parameter values obtained from the fittings. The obtained temperature-dependent profiles of $S$, for  both  the  cases,  are  shown in  Figure 4. The order parameter $S$ decreases appreciably after the incorporation of QDs. The decrease in order parameter can be ascribed to the reduction of cybotactic cluster size after QDs incorporation, as will be discussed in the dielectric studies section. The nematic phase range, as observed from the birefringence measurements, were found around 134-106 $^\circ$C for the pure LC and around 134-104 $^\circ$C for the QDs incorporated LC, complying with the POM observations.\\

\subsection{Dielectric Studies}

\begin{figure}[b]
  \centering
  \includegraphics[width = \linewidth]{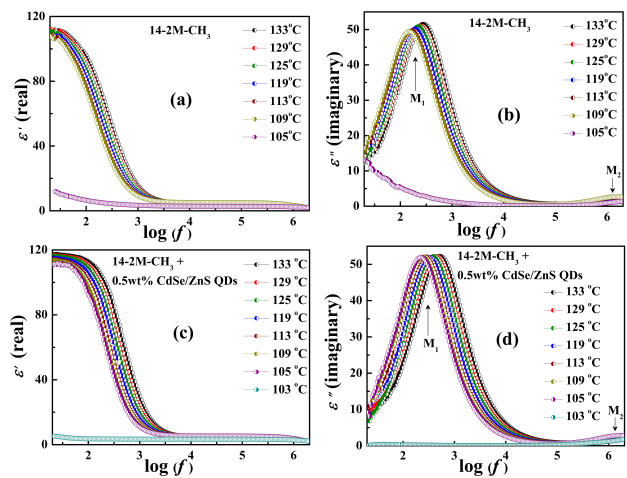}
  \caption{Frequency-dependent real ($\epsilon'$) and imaginary ($\epsilon ''$) parts of dielectric permittivity of (a-b) pristine LC (14-2M-CH$_3$) and (c-d) QDs dispersed LC nanocomposite (14-2M-CH$_{3}$ + 0.5wt\% CdSe/ZnS QDs), at different temperatures, during cooling. ($f$ in Hz)}
  \label{Fig5}
\end{figure}

Dielectric measurements have been carried out in a frequency range of 20 Hz $-$ 2 MHz (measuring voltage V$_{rms}$ = 0.2 V) and at different temperatures during the cooling cycle. The complex dielectric permittivity ($\epsilon^*$) of LCs, in the frequency domain, is expressed as, $\epsilon^*(f) = \epsilon'(f) - i\epsilon''(f)$ \cite{Haase_relaxation}. Here, $\epsilon'$ and $\epsilon''$ are the real and the imaginary parts of the complex dielectric permittivity, respectively. The dielectric spectra of $\epsilon'$ and $\epsilon''$, obtained from experiments, for the LC 14-2M-CH$_3$ and its QDs dispersed nanocomposite are shown in Figure 5. The maximum experimental error for the dielectric measurements lie within $\pm 1\%$. From Figure 5(a), we can see that the value of $\epsilon'$ at lower frequencies is $\sim$ 110, for the LC 14-2M-CH$_3$. Such high values of $\epsilon'$ have been recently observed in bent-core LCs containing cybotactic clusters \cite{Shankar_Cybo_AFM, Shankar_Cybo_CPC}. The dielectric absorption spectra unveils the associated relaxation processes in the medium. The absorption spectra of 14-2M-CH$_3$ is depicted in Figure 5(b). At any temperature, two distinct relaxation peaks (or modes) can be identified: a low-frequency mode (M$_1$) and a high-frequency mode (M$_2$). The two modes represent different relaxation processes present in the LC medium. Collective relaxation processes (due to cybotactic clusters) are known to give rise to low-frequency absorption peaks similar to M$_1$, and they are widely encountered in the N phases of bent-core LCs \cite{Haase_relaxation, Ghosh_BLC_Ferro_JMCC, Shankar_Cybo_AFM, Shankar_Cybo_CPC, Scaramuzza_BLC_JAP, Jakli_Cybo_DirectObs_PRL}. The relaxation frequencies ($f_R$) associated with cybotactic clusters can vary in the range of few tens of Hz to a few hundred Hz \cite{Shankar_Cybo_AFM, Shankar_Cybo_CPC, Ghosh_BLC_Ferro_JMCC, Scaramuzza_BLC_JAP}. Therefore, the mode M$_1$ is attributed to collective relaxation processes originating from cybotactic clusters present in the N phase of the LC. These clusters only occupy a fraction of the volume, and not all molecules form these clusters \cite{madhusudana2017two,patranabish2019one}. Experiments  show that the clusters can also exist in the isotropic phase, and their size does not change significantly across the I-N transition - a unique property of BLCs that warrants further investigation \cite{Ghosh_BLC_Ferro_JMCC, Patranabish_JML, madhusudana2017two, Panarin_Vij_Cybo_ratio_BJN, Wiant_BLC_PRE}. As reported in \cite{Patranabish_JML}, through detailed small-angle X-ray scattering (SAXS) and dielectric measurements, the N phase of the pristine LC 14-2M-CH$_3$ is cybotactic in nature, \textit{i.e.}, it contains smectic-like cybotactic clusters. Also, the mode M$_1$ is not associated with ionic impurities because no polarization response (ionic) could be detected for both the pure and the doped LCs (applied voltage up to 80 $V_{pp}$, frequencies between mHz to kHz range) \cite{Jakli_BLC_LCR, Patranabish_JML}. The high-frequency mode M$_2$ represents reorientation of the LC molecular short-axis (around the long molecular axis), subject to planar anchoring conditions \cite{Ghosh_BLC_Ferro_JMCC, Patranabish_JML}. On entering the crystalline phase, M$_2$ was no more visible, which signifies that M$_2$ is a feature of the LC phase itself and it is not related to the cell's ITO electrodes. Further, with increasing temperature, the strength of M$_2$ is decreasing. It suggests that in the isotropic phase, at temperatures much higher than the isotropic-nematic transition, the mode M$_2$ will be completely absent. Therefore, we attribute M$_2$ to the reorientation of the LC molecular short-axis. Similar high-frequency modes were observed in a 5-ring bent-core LC CNRbis12OBB in the N phase and they were attributed to the independent rotation of dipolar groups (around the long axis) \cite{tadapatri2010permittivity}.\\

\begin{figure}[b]
  \centering
  \includegraphics[width = \linewidth]{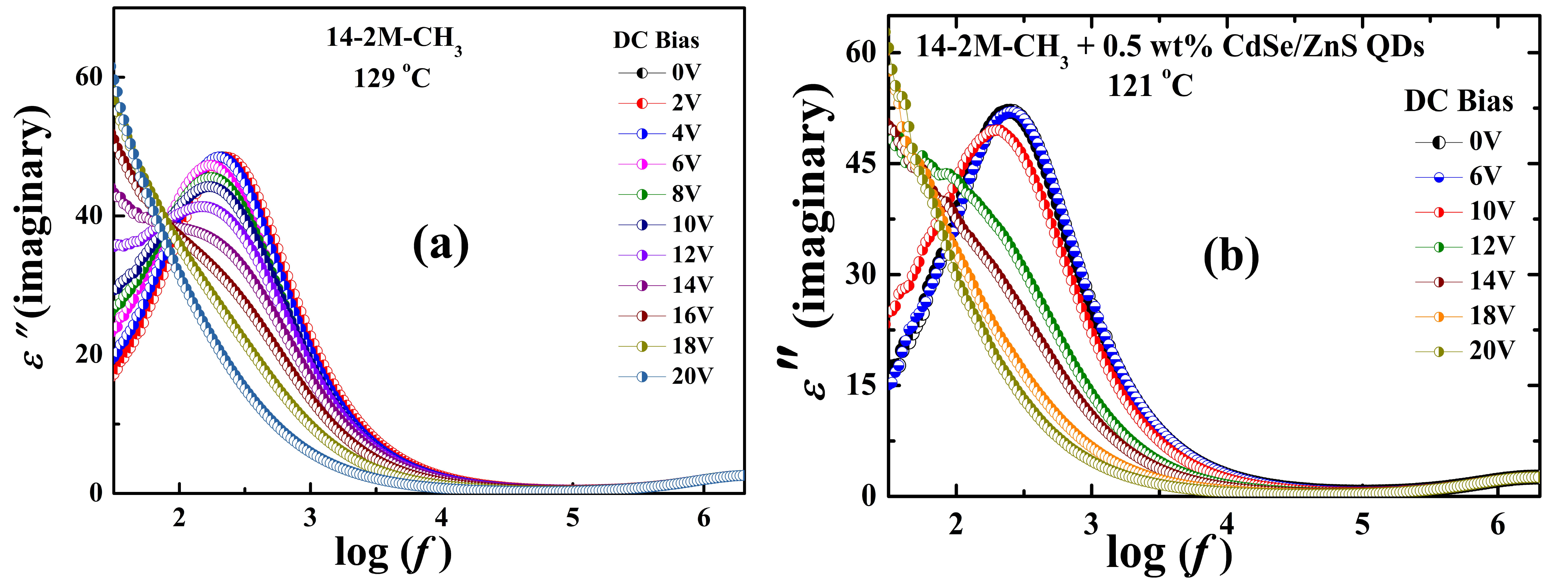}
  \caption{DC bias suppression of the low-frequency relaxation mode (M$_1$) in  (a) pure 14-2M-CH$_3$ at 129 $^\circ$C and (b) 0.5 wt\% QDs incorporated 14-2M-CH$_3$ at 121 $^\circ$C. ($f$ in Hz)}
  \label{Fig6}
\end{figure}

\begin{figure}[t]
  \centering
  \includegraphics[width = 0.6\linewidth]{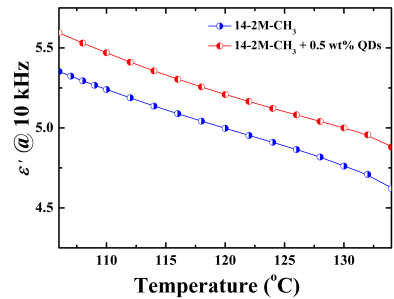}
  \caption{Dielectric permittivity of 14-2M-CH$_3$ and its nanocomposite at 10 kHz, as a function of temperature.}
  \label{Fig7}
\end{figure}

For the 0.5 wt\% QDs incorporated 14-2M-CH$_3$, the dispersion curve is shown in Figure 5(c). Similar to the pristine LC 14-2M-CH$_3$, we can see that the value of $\epsilon'$ at lower frequencies is large ($\sim$ 110). The absorption spectra of the LC nanocomposite is depicted in Figure 5(d). At any temperature, two distinct relaxation peaks (or modes) can be identified: a low-frequency mode (M$_1$) and a high-frequency mode (M$_2$). After the incorporation of QDs, a relative change in the associated relaxation frequencies ($f_R$) of the modes can be observed, compared to the pristine LC. The values of $f_R$ have been evaluated from the experimental data and it has been discussed in detail later in this section. The $f_R$ associated with M$_1$ is denoted by $f_{R1}$ and for M$_2$, it is denoted by $f_{R2}$. By comparison with the results obtained for the pristine LC 14-2M-CH$_3$, it is evident that the collective processes (and hence the cybotactic clusters) survive in the QDs dispersed LC nanocomposite. However, to establish this firmly, additional DC bias measurements have been performed. For collective processes, when a DC bias voltage is applied across a LC cell, the relaxation process ceases to exist. As a result, the dielectric relaxation modes get suppressed and at high voltages they become extinct \cite{douali_dielectric, Ghosh_BLC_Ferro_JMCC, Haase_relaxation}. A DC bias voltage of amplitude up to 20 V was applied across the LC cell and the dielectric measurements were performed (Figure 6). For the pure LC 14-2M-CH$_3$, a continuous and gradual suppression of mode M$_1$ with an applied DC bias voltage is observed (Figure 6(a)). It is a confirmatory proof of collective relaxations, and hence the presence of cybotactic clusters in the N phase of the LC \cite{Ghosh_BLC_Ferro_JMCC,Patranabish_JML}. Similarly to the pristine LC, we observe that the mode M$_1$ of the LC nanocomposite becomes suppressed (Figure 6(b)), and then completely absent at higher voltages ($\sim$ 20 V). This observation further confirms the collective behaviour of M$_1$ \cite{Ghosh_BLC_Ferro_JMCC,Patranabish_JML}, and hence corroborates retention of the cybotactic nematic phase (N$_{Cyb}$) in the QDs dispersed LC nanocomposite. The high-frequency mode M$_2$, however, does not show any change on DC bias application and hence it represents reorientation of LC molecular short axis. Moreover, we note that in the doped LC, similar to the pristine LC, M$_2$ is absent in the crystalline state and its strength decreases with increasing temperature.\\

The permittivity ($\epsilon'$) values at $f =$ 10 kHz, have been evaluated as a function of temperature (Figure 7). It shows that on incorporation of QDs, the permittivity ($\epsilon'$) increases appreciably. In a planar configuration, $\epsilon'$ represents $\epsilon_{\perp}$. The dielectric anisotropy ($\Delta \epsilon$) is defined as, $\Delta \epsilon = \epsilon_{||} - \epsilon_{\perp}$. Therefore, an increase in $\epsilon'$ implies a decrease in $\Delta \epsilon$. Further, a reduction in the dielectric anisotropy is indicative of decreasing macroscopic order parameter (since $\Delta \epsilon \propto S$) \cite{JitendraK_QDBLC_JML, maier_orderparameter}. It agrees well with the observations made in sections III-B and III-A.\\

To analyze the dielectric modes and the effects of incorporation of QDs, the associated dielectric parameters (e.g. dielectric strength ($\delta \epsilon$), relaxation frequency ($f_R$)) have been evaluated by fitting the experimental dielectric data (both $\epsilon'$ and $\epsilon''$, simultaneously), using the well-known Havriliak-Negami fit function. The frequency-dependent complex dielectric permittivity, $\epsilon^{*}(f) $, can be described by the modified Havriliak-Negami (H-N) equation, \cite{Havriliak_1966, Havriliak_1967,Ghosh_HNFit_JML,Ghosh_HNFit_LC,susantaJMC} which also includes contributions from the dc conductivity ($\sigma_0$):
\begin{equation}
\epsilon^*(f) = \epsilon_{\infty} + \sum_{k=1}^N \frac{\delta \epsilon_k}{[1 + (i 2\pi f\tau_k)^{\alpha_k}]^{\beta_k}} - \frac{i \sigma_0}{\epsilon_0 (2 \pi f)^s}	 
\end{equation}

\begin{figure}[b]
  \centering
  \includegraphics[width = \linewidth]{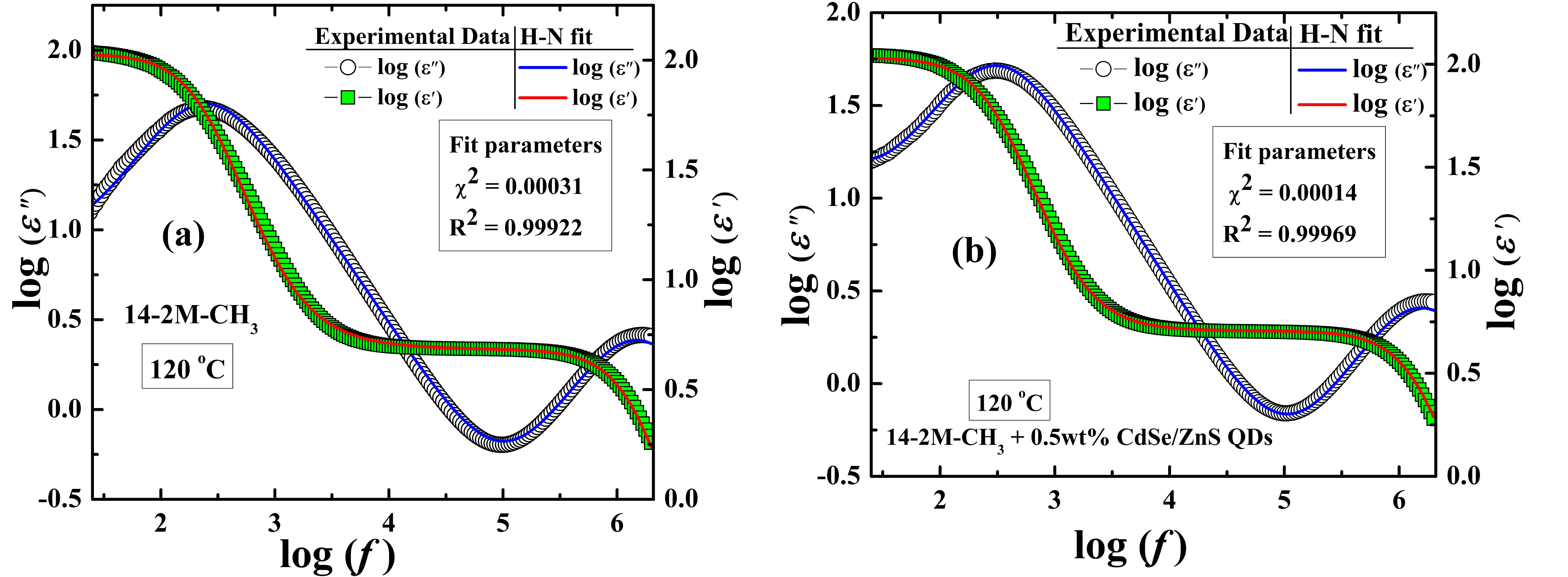}
  \caption{Simultaneous fitting of the real ($\epsilon'$) and the imaginary ($\epsilon''$) parts of complex dielectric permittivity (in \textit{log} scale) using the Havriliak-Negami (H-N) equations in - (a) pure 14-2M-CH$_3$ and (b) 0.5 wt\% QDs incorporated 14-2M-CH$_3$ ($f$ in Hz). Experimental data - The green squares represent $\epsilon'$ and the hollow circles represent $\epsilon''$. Fit data - The red solid line represents fit to $\epsilon'$ and the blue solid line represents fit to $\epsilon''$. The fitting parameters $\chi^2$ and $R^2$ are generated by the fitting algorithm, so that $\chi^2 \sim 0$ and $R^2 \sim 1$ describe good fits.}
  \label{Fig8}
\end{figure}

The last term on the right-hand side of equation (4) describes the motion of free-charge carriers in the sample. The characteristic dielectric parameters such as the relaxation frequency ($f_R$) and the dielectric strength ($\delta \epsilon$) are obtained by fitting the experimental dielectric permittivity ($\epsilon'$) and dielectric loss ($\epsilon''$) data simultaneously to the real and the imaginary parts of equation (4) given by \cite{Ghosh_BLC_Ferro_JMCC,Ghosh_HNFit_JML,Ghosh_HNFit_LC,Golam_Hockey_BLC_ACSOmega,susantaJMC},
\small{
\begin{equation}
\begin{aligned}
    & \epsilon'  = \epsilon_{\infty} + \\
    & \sum_{k=1}^N \frac{\delta \epsilon_k  \cos (\beta_k \theta)}{[1+(2\pi f \tau_k)^{2\alpha_k} + 2(2\pi f\tau_k)^{\alpha_k} \cos (\alpha_k \pi / 2)]^{\beta_k/2}} \\
    \end{aligned}
\end{equation}
}

\small{
\begin{equation}
\begin{aligned}
    & \epsilon''  = \frac{\sigma_0}{\epsilon_0 (2 \pi f)^s} + \\
    & \sum_{k=1}^N \frac{\delta \epsilon_k  \sin (\beta_k \theta)}{[1+(2\pi f \tau_k)^{2\alpha_k} + 2(2\pi f\tau_k)^{\alpha_k} \cos (\alpha_k \pi / 2)]^{\beta_k/2}} \\
    \end{aligned}
\end{equation}
}

Here, 
\begin{equation}
\theta= \tan^{-1} \left[ \frac{(2 \pi f \tau_k)^{\alpha_k}\sin(\alpha_k \pi/2)}{(1+ (2 \pi f \tau_k)^{\alpha_k}  \cos(\alpha_k \pi/2)} \right]	
\end{equation}

\begin{figure}[b]
  \centering
  \includegraphics[width = \linewidth]{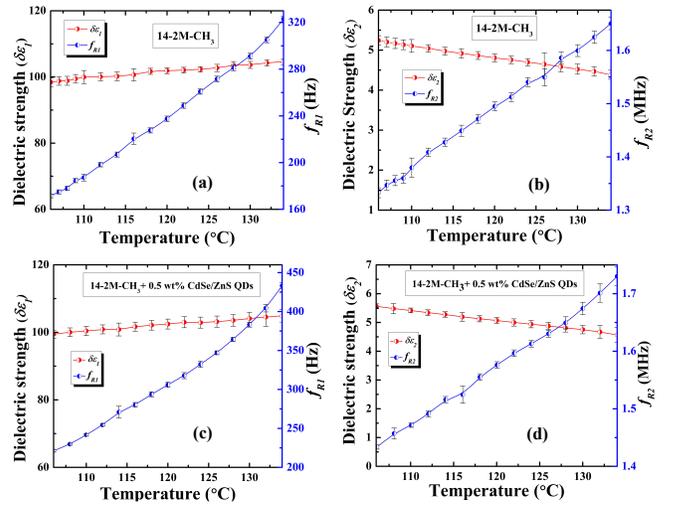}
  \caption{Temperature-dependent variation of the relaxation frequency ($f_R$) and the dielectric strength ($\delta \epsilon$) corresponding to M$_1$ and M$_2$ of (a-b) the pristine LC and (c-d) 0.5 wt $\%$ QDs incorporated LC.}
  \label{Fig9}
\end{figure}

Here, $f$ is the frequency, $\epsilon_{\infty}$ is the high-frequency limit of permittivity, $\delta \epsilon_k$ is the dielectric strength for $k$-th relaxation process, $\sigma_0$ is the dc conductivity, $\epsilon_0$ is the free-space permittivity ($8.854*10^{-12}$ F/m), $s$ is a fitting parameter responsible for the nonlinearity in dc conductivity part (for ohmic behaviour, $s$ = 1), $k$ is the number of relaxation processes, $\tau_k ( = 1/2\pi f_k)$ is the relaxation time for $k$-th relaxation process, $\alpha_k$ and $\beta_k$ are the empirical fit parameters that describe symmetric and non-symmetric broadening, respectively, of the $k$-th relaxation peak. In our case, in the absorption curve, we have two different relaxation peaks and hence $k =$ 1 and 2. A representative of the obtained Havriliak-Negami (H-N) fits are shown in Figure 8. The values of $\alpha_1$ and $\alpha_2$ lie in the range of 0.97 $-$ 1 while the values of $\beta_1$ and $\beta_2$ lie in the range of 0.93 $-$ 1 (we perform the fitting over a range of temperatures $106^{\circ}$C$-134^{\circ}$C and the ranges in $\alpha_1 ..\beta_2$ are specified). In the study of a 5-ring bent-core LC C1Pbis10BB and its mixtures with a calamitic nematic LC 6OO8, the authors reported a Debye-type low-frequency relaxation mode $B_{||1}$ \cite{salamon2010dielectric}. They also write that smectic-like clusters can induce a Debye-type relaxation in the low-frequency region of dielectric spectrum. Our dielectric results also indicate that M$_1$ is a near Debye-like relaxation process and the associated relaxation frequencies overlap with the mode $B_{||1}$ reported in \cite{salamon2010dielectric}. The variations of the relaxation frequency ($f_R$) and the dielectric strength ($\delta \epsilon$) of modes M$_1$ and M$_2$ with temperature, as obtained from the fitting, are shown in Figure 9. The results show that $\delta \epsilon_1$ ($i.e.$ corresponding to M$_1$) for 14-2M-CH$_3$ increases slightly, from $\sim$ 98 to $\sim$ 104, with increasing temperature. Similarly, $\delta \epsilon_1$ for the QDs dispersed nanocomposite increases slightly, from $\sim$ 100 to $\sim$ 105, with increasing temperature. Thus, the dielectric strength $\delta \epsilon_1$ is largely unaffected on doping. Again, the value of $\delta \epsilon_1$ is quite large and it is similar to other bent-core LCs with cybotactic clusters \cite{Shankar_Cybo_AFM, Shankar_Cybo_CPC, Ghosh_BLC_Ferro_JMCC}. The dielectric strength $\delta \epsilon_2$ associated with M$_2$ is found to be very small and increases with decreasing temperature - from $\sim$ 4.5 to 5.5 for both 14-2M-CH$_3$ and its nanocomposite.\\

The relaxation frequency ($f_{R1}$) associated with M$_1$ lies in the range of $\sim$ 170 Hz to $\sim$ 320 Hz for the pure LC, similar to several other bent-core LCs with cybotactic clusters \cite{Shankar_Cybo_AFM, Shankar_Cybo_CPC, Ghosh_BLC_Ferro_JMCC}. The incorporation of QDs causes $f_{R1}$ to shift to higher frequencies ($\sim$ 220 Hz to $\sim$ 430 Hz). It indicates that there is an apparent reduction in size of the smectic-like cybotactic clusters \cite{Panarin_Vij_Cybo_ratio_BJN}. This reduction can be estimated qualitatively by taking the ratio of the relaxation frequency $f_{R1}$, for the pristine LC and its doped counterpart. The ratio has an average value $\sim$ 0.67, where the average is taken over two sets of measurements and a range of temperatures. This ratio signifies a relative change in the average number of molecules ($N_c$) present in each cluster (in accordance with our earlier theoretical model and the experiments) \cite{patranabish2019one,Panarin_Vij_Cybo_ratio_BJN}. The decrease in the measured order parameter ($S$) on doping can be ascribed to  reduced cluster sizes on QDs incorporation. In our earlier theoretical work on bent-core nematic LCs, we take $N_c$ = 50 \cite{patranabish2019one}. Recent observations have shown that the typical size of smectic-like cybotactic clusters lie in the range of few tens of nanometres to around a hundred nanometres \cite{Jakli_Cybo_DirectObs_PRL}. Again, the typical dimension of a bent-core LC molecule is around 2$-$3 nanometres. Therefore, the number $N_c$ = 50 is justified in the case of pure (undoped) bent-core LCs, with cybotactic clusters. For the QDs dispersed bent-core nematic LCs, we can take $N_c$ $\sim$ 33 (= 50 x 0.67) as a reasonable value. The relaxation frequency $f_{R1}$ manifests a gradual decrease with decreasing temperature, divulging an Arrhenius-type behaviour ($f_R=f_0 \exp(-E_a/k_B T)$; $f_0$ is a temperature-independent constant, $E_a$ is the activation energy, $k_B$ is the Boltzmann’s constant and $T$ is the absolute temperature). $f_{R2}$ also  demonstrates an Arrhenius-like behaviour. \\

\begin{figure}[t]
  \centering
  \includegraphics[width =\linewidth] {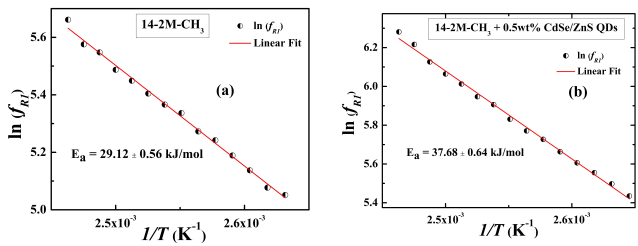}
  \caption{Arrhenius plot of the M$_1$ relaxation frequency ($f_{R1}$) in the nematic (N) phase of (a) pristine and (b) 0.5wt\% CdSe/ZnS QDs incorporated 14-2M-CH$_3$. The activation energy ($E_a$) is calculated from the slope of the linear fit, represented by the solid red line.}
  \label{Fig10}
\end{figure}

The activation energy ($E_a$) associated with a relaxation process encodes the minimum amount of energy required for that process to take place \cite{Haase_relaxation}. The value of $E_a$ associated with the relaxation processes can be obtained by plotting $f_R$ as a function of $1/T$, using the relation $f_R=f_0 \exp(-E_a/k_B T)$. The Arrhenius plots of $ln$($f_{R1}$) vs. $1/T$ (for M$_1$) for the two compounds are shown in Figure 10. The activation energy ($E_a$), associated with M$_1$, is
evaluated from the slope of the linear fit, as shown in Figure 10. The value of $E_a$ ($\sim$ 29.12 kJ/mol) increases significantly after the incorporation of QDs ($\sim$ 37.68 kJ/mol). For a small cluster size, the cluster's dipole moment ($\mu$) also becomes small. Hence, more energy is required to interact with an external electric field.  Therefore, an increased value of $E_a$ for M$_1$, after the incorporation of CdSe/ZnS QDs, implies a decrease in the size of cybotactic clusters. It also concurs with our earlier observations. The activation energy associated with M$_2$ has rather small values ($\sim$ 8 kJ/mol) and it does not change significantly after the incorporation of QDs.

\section{Mathematical Model}

In this section, we  propose a simple mathematical model that describes the QD doping-induced reduction of the nematic scalar order parameter for dilute suspensions, that could  subsequently be improved to describe novel features such as ferroelectricity, chirality, biaxiality and transition pathways between multiple stable states. Since experimental domain is a simple planar cell, denoted by $\Omega \subset \mathbb{R}^3$, with height about 5 microns ($5 \times 10^{-6} \mathrm{m}$), we assume a characteristic length of the system 
 \begin{equation}
   x_s = 5 \times 10^{-8} \mathrm{m},
 \end{equation}
 as in \cite{patranabish2019one}, so that the cell thickness is $100$ units of $x_s$. The cross-sectional dimensions of the cell are much larger than the cell height, so it is reasonable to assume that structural variations only occur across the cell height i.e. this is a one-dimensional problem. We assume that the QDs are spherical in shape, with an average radius of $2.8$ nanometres; the size of the QDs is much smaller than the typical separation between them and the total volume occupied by the QDs is small. Let  $R$ denote the radius of a QD ($2.8$ nanometres as reported in these experiments) and we define a small parameter $\epsilon$ so that  
 $$ \epsilon^{\alpha} = \frac{R}{x_s} = R_0 = 0.056$$ for some $1<\alpha < \frac{3}{2}$. The definition of $\epsilon$ is not unique, provided it is a small parameter, relevant for \emph{dilute uniform suspensions of QDs} \cite{canevari2019design}. In particular, our mathematical model is \emph{restricted} to dilute suspensions and will need to be modified for non-dilute systems.

In \cite{madhusudana2017two}, N. V. Madhusudana proposes a Landau-de Gennes (LdG) type two-state model for the N phase of BLCs, accounting for cybotactic clusters. This two-state model is a phenomenological model based on the premise that the N phase of the BLC comprises two different types of molecules: ground state (GS) molecules and excited state (ES) molecules. The ES molecules define the smectic-like cybotactic clusters and the GS molecules are located outside the clusters. The generic LdG theory models a single component system e.g. the GS molecules, typically assumed to be rod-like molecules that tend to align with each other, yielding long-range orientational order \cite{ravnik2009landau}. Madhusudana's model is a two component model, the GS and ES molecules, with additional coupling effects and in \cite{patranabish2019one}, we describe the N phase of the BLC by two macroscopic tensor order parameters (with a number of simplifying assumptions)
\begin{equation}
\begin{aligned}
\Q_g =  \sqrt{\frac{3}{2} }  S_g \left( \n_g \otimes \n_g - \frac{1}{3} \mathrm{I} \right)\\
\quad \Q_c =  \sqrt{\frac{3}{2} }  S_c \left( \n_c \otimes \n_c - \frac{1}{3} \mathrm{I} \right)
\end{aligned}
\end{equation}
respectively, where $\Q_g$ is the LdG order parameter for the GS molecules, and $\Q_c$ is the LdG order parameter associated with the ES molecules. In both cases, we assume that $\Q_g$ and $\Q_c$ have uniaxial symmetry i.e. the GS (ES) molecules align along a single averaged distinguished direction $\n_g$ (respectively $\n_c$), and assume that $\n_g$ and $\n_c$ are constant unit-vectors or directions, \textbf{so that there are no director distortions or defects}. There are two scalar order parameters, $S_c$ and $S_g$, corresponding to ES (excited state) and GS (ground state) molecules respectively. As is standard with variational approaches, the experimentally observed configurations are described by local or global minimizers of a suitably defined LdG-type energy in terms of $S_c$, $S_g$ and the coupling between them. 

In \citeauthor{patranabish2019one}, the authors theoretically study stable $(S_g, S_c)$ profiles in a simple planar cell geometry (as experimentally studied in this manuscript), as a function of the temperature, in terms of minimizers of the following LdG-type energy (heavily inspired by the work in \cite{madhusudana2017two} with additional elastic effects that account for spatial inhomogeneities):
\begin{equation}\label{Energy1}
 \begin{aligned}
   \mathcal{F} & =  \int_{\Omega} (1 - a_x) \left( \frac{a_g}{2}(T - T^{*})S_g^2 - \frac{B_g}{3} S_g^3 + \frac{C_g}{4} S_g^4 - E_{el} S_g    \right) \\
   & + \frac{a_x}{N_c} \left( - (1 - a_x) \gamma S_g S_c + \frac{\alpha_c}{2} S_c^2 + \frac{\beta_c}{4} S_c^4  \right) \\
   &- a_xJ E_{el} S_c  + K_g |\nabla S_g|^2 + K_c |\nabla S_c|^2 \dd \x  \\
   \end{aligned}
\end{equation}
Here, the subscripts $g$ and $c$ denote the GS and the
ES molecules, respectively, and the clusters are essentially
formed by the ES molecules. They work with a one-constant approximation and $K_g$ and $K_c$ are the elastic
constants of the GS and ES molecules respectively. $a_g$,$B_g$,$C_g$ are the material dependent parameters in the LdG free energy, $T^*$ is the nematic supercooling temperature such that the isotropic phase of the GS phase is unstable for $T<T^*$.
The parameter, $\gamma$, is the coupling parameter between the GS molecules and the
clusters \cite{madhusudana2017two}. The coefficients, $\alpha_c$ and $\beta_c$, are saturation parameters to ensure that the absolute value of $S_c < 1$ for physically relevant parameter regimes , $N_c$ is the number of ES molecules in each
cluster, $a_x$ is the mole fraction of the ES
molecules. $J$ accounts for the shape anisotropy of ES molecules.
$E_{el}$ is the electric field energy ($\frac{1}{2}\epsilon_0 \Delta \epsilon E^2$) where $\epsilon_0$ is the
free-space permittivity, $\Delta \epsilon$ is the dielectric anisotropy, $E$ is
the applied electric field. 

The mathematically and physically pertinent question is - how is the energy (\ref{Energy1}) modified by the uniformly suspended QDs in the dilute limit? Following the elegant homogenisation framework developed in \cite{canevari2019design}, the additional doping can be described by an effective field in the dilute limit, and \textbf{this effective field strongly depends on the shape and anchoring conditions on the QDs, but not the size of the QDs in the dilute limit (as will be made clearer below and the size will matter in the non-dilute limit).}
We assume the QDs are spherical in shape, as stated above, and impose preferred alignment tensors on the QD surfaces, $\Q_v^g$ ( $\Q_v^c$), for QDs outside (inside the) clusters respectively.
We assume that $\Q_v^g$ and $\Q_v^c$ are constant tensors, given by,
\begin{equation}
  \Q_v^g =  \sqrt{\frac{3}{2} }  S_g^b \left( \n_g \otimes \n_g - \frac{1}{3} \mathrm{I} \right).
\end{equation}
and
\begin{equation}
  \Q_v^c =  \sqrt{\frac{3}{2} }  S_c^b \left( \n_c \otimes \n_c - \frac{1}{3} \mathrm{I} \right).
\end{equation}
for some fixed $S_g^b, S_c^b > 0$. There is no clear argument for the choice of $S_g^b$ and $S_c^b$ in this model but we make reasonable choices below. \textbf{Further, we assume that $\n_g$ ($\n_c$) is the same for $\Q_g$ and $\Q_v^g$ (likewise for $\Q_c$ and $\Q_v^c$), so that there is no director distortion at the QD interfaces.} Assuming a Rapini-Papoular surface anchoring energy on the QD surfaces, the QD surface energy density is given by
\begin{equation}\label{surface_reduced}
f_s^g(S_g, S_c, S_g^b, S_c^b) = W_g^0  |S_g - S_g^b|^2 + W_c^0  |S_c - S_c^b|^2,
\end{equation}
where $W_g^0$ and $W_c^0$ are the anchoring coefficients on the QD-GS interfaces, QD-ES interfaces respectively.
For relatively strong anchoring on the QD interfaces,  $W_c^0$ and $W_g^0$ can be taken to be approximately $1 \times 10^{-2} \mathrm{J/m}^2$  \cite{ravnik2009landau}. In particular, $W_c^0$ is zero for QDs outside the clusters and $W_g^0$ is zero for QDs inside the clusters. Next, we follow the paradigm in \cite{canevari2019design} to describe the collective effects of a uniform dilute suspension of QDs, with surface energies as in (\ref{surface_reduced}), in terms of an additional \emph{homogenised} effective term in the free energy (\ref{Energy1}).

As in \cite{patranabish2019one}, we let $A = (1 - a_x) a_g (T - T^{*}),~~ B =  (1 - a_x) B_g,~~C = (1 - a_{x}) C_g, D = a_x(1 - a_x) \gamma / N_c, E = (1 - a_x) E_{el}, \quad M = \alpha_c a_{x} / N_c, N = \beta_c a_x / N_c, \quad P = J E_{el} a_x, \quad W_g = (1 - a_x) W_g^0$ and $W_c = \frac{a_x}{N_c} W_c^0$,  
 where $a_x$ is the fixed mole fraction of the ES molecules. Moreover, we assume the $E_{el} = 0$ throughout this paper. In agreement with the parameter values used in \cite{patranabish2019one}, we use fixed values $K_g =
K_c = K = 15pN$ ; $a_g = 0.04, B_g = 1.7,C_g = 4.5,\alpha_c = 0.22, \beta_c = 4.0$ ($\alpha_g,B_g,C_g,\alpha_c$
and $\beta_c$ in $10^6/4$ SI units). 
Following the methods in \cite{canevari2019design}, we describe the dilute QD-doped BLC system by means of the  total free energy below, without rigorous justification but rather as a phenomenological model to describe salient features of novel nanocomposites.
 \begin{equation}\label{Energy2}
   \begin{aligned}
   \mathcal{F} & =  \int \left( \frac{A}{2}S_g^2 - \frac{B}{3} S_g^3 + \frac{C}{4} S_g^4 \right) \\
   & +  \left( - D S_g S_c + \frac{M}{2} S_c^2 + \frac{N}{4} S_c^4  \right)  + K_g |\nabla S_g|^2 + K_c |\nabla S_c|^2 \dd \x  \\
   & +   \epsilon^{3 - 2\alpha}\int_{\pp \mathcal{P}} W_g |S_g - S_g^b|^2 \dd S + \epsilon^{3 - 2\alpha}\int_{\pp \mathcal{P}}  W_c |S_c - S_c^b|^2 \dd S, \\
   \end{aligned}
 \end{equation}
\\ where $\mathcal{P}$ is the collection of the QDs in the suspension and $1 < \alpha< \frac{3}{2}$, so that $\epsilon^{ 3- 2\alpha} \to 0$ as $\epsilon \to 0$. The pre-factor of $\epsilon^{3 - 2\alpha}$ is specific to dilute systems. \textbf{The main novelty is the surface energy term originating from the QD-GS and QD-ES interfaces, and the homogenized effective field is derived in the $\epsilon \to 0$ limit, as will be discussed below.}
 
 We non-dimensionalize the free-energy (\ref{Energy2}) by letting 
$\bar{\x} = \x/x_s, \quad \bar{S_g} = \sqrt{\frac{27C^2}{12 B^2}}S_g, \quad \bar{S_c} = \sqrt{\frac{27 C^2}{12 B^2}}S_c \quad  \bar{\mathcal{F}} = \frac{27^2 C^3}{72 B^4 x_s^3}\mathcal{F}$,
Dropping all \emph{bars} for convenience (so that $S_g$ and $S_b$ denote the scaled order parameters), the dimensionless energy is (we take $E_{el} = 0$)

\begin{equation}   
  \begin{aligned} \label{EnergyH}
   \mathcal{F} & =  \int_{\Omega_\epsilon} \left( \frac{t}{2}S_g^2 - S_g^3 + \frac{1}{2} S_g^4  \right) +  \left( - C_1 S_g S_c + C_2 S_c^2 + C_3 S_c^4  \right)  \\
   & + \kappa_g \left(\frac{d S_g}{dx} \right)^2 + \kappa_c \left(\frac{d S_c}{dx} \right)^2 \dd \x  \\
   & +  \epsilon^{3 - 2\alpha}\int_{\pp \mathcal{P}}  w_g |S_g - S_g^b|^2 \dd S + \epsilon^{3 - 2 \alpha} \int_{\pp \mathcal{P}} w_c |S_c - S_c^b|^2 \dd S, \\
  \end{aligned}
\end{equation}
where $\Omega_\epsilon$ is the three-dimensional planar cell with the QDs removed, $\mathcal{P}$ is the collection of the three-dimensional spherical QDs with re-scaled radius $\epsilon^{\alpha}$ for $1 < \alpha < \frac{3}{2}$ (see the definition of $\epsilon$ above), and
\begin{equation}
  \begin{aligned}
    & t = \frac{27 AC}{6 B^2}, \quad C_1  = \frac{27 CD}{6 B^2}, \quad C_2 = \frac{27 C M}{12 B^2}, \quad C_3 = \frac{N}{2C},\\
    & \kappa_g = \frac{27 C K_g }{6 B^2 x_s^2}, \quad \kappa_c = \frac{27 C K_c}{6 B^2 x_s^2} \\
    & w_g = \frac{27 C W_g}{6 B^2 x_s}  \quad w_c = \frac{27 C W_c}{6 B^2 x_s} . \\
  \end{aligned}
\end{equation} Note that $|\nabla S_g|^2 = \left(\frac{d S_g}{dx} \right)^2$ since we assume that structural variations in $S_g$ and $S_c$ only occur across the cell height, $0\leq x \leq 100$ (recall the choice of $x_s$ above). 


In \cite{canevari2019design}, the authors study minimizers of free energies of the form, 
\begin{equation}
    \label{eq:homnew}\int\int\int_{\Omega_\epsilon} f_{el}(\nabla \Q ) + f_b (\Q) dV + 
    \epsilon^{3 - 2\alpha} \int\int_{\pp \mathcal{P}} f_s\left(\Q, \nu \right) dA,  \end{equation} with $1 < \alpha < \frac{3}{2}$, where $f_{el}(\nabla \Q )$ is a general convex function of the gradient of an arbitrary order parameter $\Q$, $f_b$ is a polynomial function of the scalar invariants of $\Q$, and $f_s$ are arbitrary surface energies on the QD interfaces. The dilute limit is described by the $\epsilon \to 0$ limit, and minimizers of (\ref{eq:homnew}) converge to minimizers of the following homogenized energy, as $\epsilon \to 0$,
    \begin{equation}
        \label{eq:homnew2}
        \mathcal{F}_h (\Q) = \int_{\Omega} f_{el}(\nabla \Q) + f_b(\Q) + f_{hom}(\Q) dV,
    \end{equation}
    where $f_{hom} = \int_{\partial \omega} f_s\left(\Q, \nu \right) dS$, $\omega$ is a representative QD and $\nu$ is the outward normal to $\partial \omega$. \textbf{In particular, the shape, anchoring conditions, material properties including encapsulation properties of the QD inclusions are absorbed in the definition of $f_{hom}$. The distortion effects around the QDs are also described by $f_{hom}$ for dilute systems.} In our case, the QDs are spherical inclusions and applying the results in \cite{canevari2019design}, we have $f_{hom} = \int_{\pp B(\mathbf{0}, 1)} f_s(\Q, \nu) dA$, $B(\mathbf{0}, 1) \subset \mathbb{R}^3$ is a generic three-dimensional unit ball and $f_s$ is the surface energy (\ref{surface_reduced}).
We apply this result to calculate the homogenized potential corresponding to (\ref{surface_reduced}), see below:
\begin{equation}
\label{eq:fhom}
  f_{hom}(S_g, S_c ) = w_{g}^{(1)} S_g^2 - w_{g}^{(2)} S_g +w_{c}^{(1)} S_c^2 - w_{c}^{(2)} S_c
\end{equation}
where
\begin{equation}
  \omega_{g}^{(1)} = 4 \pi w_g, \quad   \omega_{g}^{(2)} = 8 \pi S_g^b w_g
\end{equation}
and 
\begin{equation}
 w_{c}^{(1)} = 4 \pi w_c, \quad   w_{c}^{(2)} = 8 \pi  S_c^b w_c.
\end{equation}

Hence, the total non-dimensionalized \emph{homogenized} free energy is given by
\begin{equation}\label{bulk_energy}
  \begin{aligned}
    \mathcal{F} =  \int_{\Omega} & \left( \left( \frac{t}{2} + w_g^{(1)} \right)S_g^2 - \sqrt{6} S_g^3 + \frac{1}{2} S_g^4  \right) \\
    & +  \left( - C_1 S_g S_c + (C_2 + w_{c}^{(1)} ) S_c^2 + C_3 S_c^4  \right)  \\
    & + \kappa_g \left(\frac{d S_g}{dx}\right)^2 + \kappa_c \left( \frac{d S_c}{dx} \right)^2  - w_{g}^{(2)} S_g  - w_{c}^{(2)} S_c  \dd \x.  \\
   \end{aligned}
\end{equation} 
For the parameter values as stated before, we have $C_1 = 0.0700692$, $C_2 = 0.0017$ and $C_3 = 0.0040$.

Then the equilibrium/ physically observable $(S_g, S_c)$ profiles are solutions of the Euler-Lagrange equations corresponding to (\ref{bulk_energy}). 
\begin{equation}\label{EL_bulk}
\begin{cases}
&  \kappa_g \frac{d^2 S_g}{dx^2} = 2 S_g^3 - 3 \sqrt{6} S_g^2 + (t + w_g^{(1)}) S_g - C_1 S_c - w_g^{(1)}  \\
& \kappa_c \frac{d^2 S_c}{dx^2} = 4 C_3 S_c^3 + (2 C_2 + 2 w_c^{(1)}) S_c  - C_1 S_g - w_c^{(2)} .\\
\end{cases}
\end{equation} These equations need to be complemented by boundary conditions for $S_g$ and $S_c$, we fix Dirichlet boundary conditions for the scalar order parameters on the bottom ($x=0$) and top ($x=100$) of the planar cell i.e.
\begin{equation} \label{dirichletbcs}
S_g = \frac{3 + \sqrt{9 - 8t}}{4}, \quad S_c = 0 ~ \textrm{on $x=0$ and $x=100$,}
\end{equation} which corresponds to the absence of clusters on the planar cell boundaries. The boundary conditions (\ref{dirichletbcs}) are not special and we believe that our qualitative conclusions would hold for other choices of the Dirichlet boundary conditions too. We assume that $\mathbf{n}_g$ and $\mathbf{n}_c$ are constant unit vectors, and our analysis is independent of the choice of $\mathbf{n}_g$ and $\mathbf{n}_c$, provided they are constant vectors. We also need to specify $S_g^b$ and $S_c^b$ to determine $\omega_g^{(2)}$ and $w_c^{(2)}$ above and we choose
\begin{equation}
S_g^b = \frac{3  + \sqrt{9 - 8t}}{4}, \quad S_c^b = 0,
\end{equation}
with $W_g = W_c = W$. 

Next, we numerically solve the coupled equations in (\ref{EL_bulk}) to compute the equilibrium profiles of $(S_g, S_c)$ as a function of temperature,  and different values of $W$. The parameters $N_c$ and $\gamma$ are coupled i.e. larger clusters are likely to have larger values of $N_c$ and $\gamma$, and we expect $N_c$ and $\gamma$ to be smaller for the doped system compared to its undoped counterpart, based on the experimental results that suggest smaller cybotactic clusters in QD-doped BLCs compared to their undoped counterparts. We define the bulk mean order parameter $S_m$, which is a weighted scalar order parameter as shown below
\begin{equation}
S_m = (1 - a_x) S_g + a_x S_c.
\end{equation}
\textbf{The weighted scalar order parameter, $S_m$, is the theoretical analogue of the measured order parameters from experimental birefringence measurements.} We use the value of $S_m$ at room temperature (293 K) with $(N_c, W, \gamma) = (50, 0, 5)$ to normalize $S_m$. Recall that $N_c=50$ and $\gamma = 5$ have been used to study the undoped BLC system in \cite{patranabish2019one}. 

\begin{figure}[!h]
  \centering
  \includegraphics[width = \linewidth]{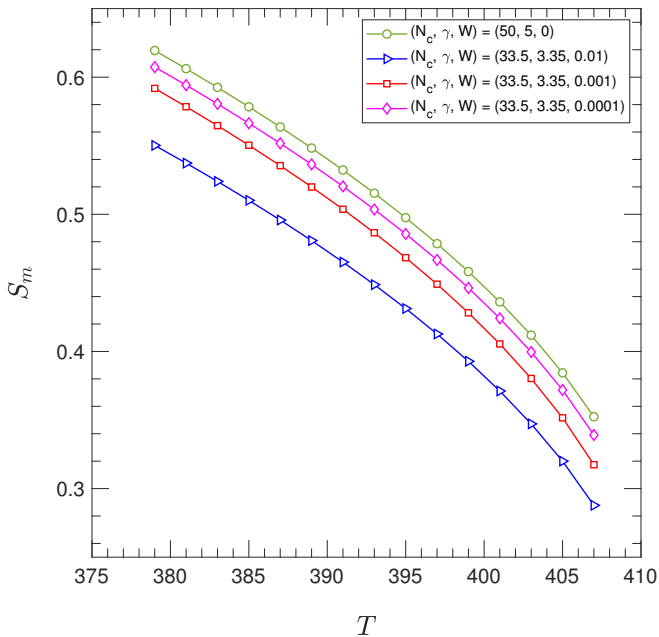}
  \caption{Bulk mean order parameter as a function of temperature for the undoped (green circle) and the doped (red square, blue triangle, purple diamond) cases (Temperature `$T$' is in K).}\label{Sm_T}
\end{figure}

In Figure \ref{Sm_T}, we plot $S_m$ as function of temperature for undoped and doped systems, for different values of $W$. For the undoped system, $(N_c, \gamma, W) = (50, 5, 0)$ by analogy with the values used in \cite{patranabish2019one}. \textbf{For QD-doped systems, the experimental results suggest that the clusters are shrunk by a factor of $0.67$ (qualitatively deduced by the ratio of the relaxation frequencies) and hence we take $N_c = 50 \times 0.67 =33.5$ and $\gamma = 5 \times 0.67 =3.35$ for doped systems.} We plot the solution, $S_m$-profiles for three doped systems -   $(N_c, \gamma, W) = (33.5, 3.35, 0.01)$,  $(N_c, \gamma, W) = (33.5, 3.35, 0.001)$, and $(N_c, \gamma, W) = (33.5, 3.35, 0.0001)$ in Figure \ref{Sm_T}.

It is clear that this simple model clearly captures the doping-induced reduction in the values of $S_m$, consistent with the experimental results in Figure~$4$. $S_m$ also decreases with increasing $T$ as expected. The numerically computed values of $S_m$ for the doped systems in Figure~\ref{Sm_T} are lower than the experimentally reported values in Figure~$4$ and we think an exact fitting between the numerical and the experimental results is unlikely at this stage. Further, the numerical results are very sensitive to the values of the anchoring coefficient and we simply do not have reliable estimates of the anchoring coefficients for QDs. In fact, the numerically computed $S_m$'s, if fitted to the experimental data, could provide an ingenious method for estimating the surface energy coefficients for QD-doped BLC systems. It is clear that the predictions of the doped model approach the predictions of the undoped model as $W \to 0$, as expected. We do not pursue this further in this manuscript.

This simple model  provides a clear explanation of why the QD-doping reduces $S_m$ in experiments - the interaction/anchoring between the QDs and the host BLC matrix increases the effective temperature (the coefficient of $S_c^2$ and $S_g^2$ in (\ref{bulk_energy})) of both the GS state and the cybotactic clusters (ES state), and hence at a given temperature, the QD-doped system experiences a shifted higher temperature and the shift is directly determined by the homogenized potential, $f_{hom}$, which in turn is determined by the anchoring and shape of the dispersed QDs. This QD-induced effective temperature shift necessarily reduces $S_m$ compared to its undoped counterpart. A doping-induced reduced $S_m$ qualitatively explains the experimentally observed reduced dielectric anisotropy and birefringence in doped systems, compared to undoped systems. \textbf{However, several open questions remain, particularly as to how we ascribe values to the cluster parameters, $N_c$ and $\gamma$, and how to describe the effects of QD-doping on these cluster parameters. Nevertheless, this is the first attempt to mathematically model dilute suspensions of QDs in the nematic phase of BLC materials, with cybotactic clusters, providing qualitative agreement with experiments.}

\section{Conclusion}
We perform experimental and theoretical studies of a QDs-dispersed BLC 14-2M-CH$_{3}$ inside a planar cell. We believe that QDs are attractive nano-inclusions for stable suspensions, without aggregation effects. We present experimental optical profiles for the pristine LC and the QDs incorporated LC nanocomposite systems, tracking the textural colour changes with temperature. We perform experimental measurements of the dielectric permittivity, including the dielectric dispersion and absorption spectra, and use fitting algorithms to calculate relaxation frequencies and dielectric strengths, that are used to validate the existence of cybotactic clusters in the doped and undoped systems, the reduction of cluster sizes in doped systems and corresponding increase in activation energies. We also present experimental measurements of the optical birefringence and the orientational order parameters of the doped and undoped systems. All the experiments demonstrate doping-induced reduction of orientational order and cluster sizes, that manifest in doping-induced reduced birefringence and a reduced dielectric anisotropy (qualitatively) at a fixed temperature. In terms of future experiments, we would like to investigate biaxiality in these QDs-dispersed BLC systems, chirality and prototype devices based on such simple planar cell geometries. For example, we could treat the planar cell to have conflicting boundary conditions on both the cell surfaces, naturally inducing inhomogeneous director profiles. 

We support some of our experimental findings with a homogenized Landau-de Gennes type model for a doped BLC system, with two scalar order parameters, $S_g$ and $S_c$, and constant director profiles. In particular, we capture the doping-induced reduction in the mean scalar order parameter which is an informative and illuminating first step. The theory can be embellished in many ways, to make it physically realistic e.g. elastic anisotropy involving additional terms in the elastic energy density, non-constant director profiles captured by non-constant $\n_g$ and $\n_c$ and understanding how the QDs affect the cybotactic clusters. This could be done by using a general two-tensor model, $\Q_g$ and $\Q_c$, without making any additional assumptions about uniaxial symmetry or constant directors, as in our mathematical model in this paper. However, it will be a challenge to describe the cybotactic cluster-mediated coupling between $\Q_g$ and $\Q_c$ without these restrictive assumptions, and some of these directions will be pursued in future work.

\section*{Credit taxonomy}
S. Patranabish and A. Sinha conducted the experiments and analysed the experimental results. Y. Wang and A. Majumdar performed the modelling and comparisons between experiments and modelling.

\section*{Acknowledgement}
The authors would like to thank Prof. N.V.S. Rao, Department of Chemistry, Assam University, Assam, India and Dr. Golam Mohiuddin, Xi'an Jiaotong University, Xi'an, China for providing the liquid crystal samples. The authors thank Dr. Giacomo Canevari for helpful discussions on the homogenised potentials. The authors also thank Anjali Sharma and Ingo Dierking for useful feedback on the experimental sections. S.P. thanks Dr. Susanta Chakraborty for useful discussions on fitting of the experimental data. S.P. acknowledges IIT Delhi for financial support under Full-time Institute Assistantship. The authors would like to thank DST-UKIERI for generous funding to support the 3-year collaborative
project.

\providecommand{\noopsort}[1]{}\providecommand{\singleletter}[1]{#1}%

\end{document}